\documentclass[review]{elsarticle}

\usepackage{lineno,hyperref}
\modulolinenumbers[5]

\usepackage{subcaption}
\usepackage{multirow}
\usepackage{footmisc}

\usepackage[T1]{fontenc}
\usepackage{caption}
\usepackage{booktabs,multirow}
\usepackage{siunitx}
\usepackage[flushleft]{threeparttable}

\journal{Journal of \LaTeX\ Templates}

\def\CAMOO{${}^{40}$Ca$^{100}$MoO$_{4}$}
\def\a{$\alpha$}
\def\b{$\beta$}
\def\g{$\gamma$}
\def\sgm{$\sigma$}
\def\cmsq{cm${}^{2}$}

\def\tm{$\times$}
\def\sm{$\sim$}
\def\lt{$<$}
\def\UNIT{ckky}

\def\znbb{$0\nu\beta\beta$}
\def\twonubb{$2\nu\beta\beta$}

\newcommand{\minus}{\scalebox{0.5}[1.0]{$-$}}
\newcommand\tmE[1]{\ensuremath{\times 10^{#1}}}
\newcommand\tmEm[1]{\ensuremath{\times 10^{\minus#1}}}

\newcommand\Em[1]{\ensuremath{10^{\minus#1}}}

\newcommand\geniso[2]{\ensuremath{\rm ^{#2}#1}}

\newcommand\newiso[3][]{
\ifx&#1&
  \expandafter\newcommand\csname #2\endcsname[1][#3]{\geniso{#2}{##1}}
\else
  \expandafter\newcommand\csname #1\endcsname[1][#3]{\geniso{#2}{##1}}
\fi
}

\newiso{K}{40}
\newiso{Th}{232}
\newiso{U}{238}
\newiso{Rn}{222}
\newiso{Ra}{226}
\newiso{Pb}{208}
\newiso{Tl}{208}
\newiso{Bi}{214}
\newiso{Ac}{226}
\newiso{Po}{210}
\newiso{Mo}{100}

\begin{document}

\begin{frontmatter}

\title{Simulations of background sources in AMoRE-I experiment}

\author[knu]{A.~Luqman}
\author[knu]{D.H.~Ha}
\author[cau]{J.J.~Lee}
\author[ibs]{E.J.~Jeon\corref{cor1}}
\cortext[cor1]{Corresponding authors}
\ead{ejjeon@ibs.re.kr}
\author[ibs]{H.S.~Jo}
\author[knu]{H.J.~Kim}
\author[ibs]{Y.D.~Kim}
\author[ibs,kriss]{Y.H.~Kim}
\author[inr]{V.V.~Kobychev}
\author[ibs]{H.S.~Lee}
\author[ibs]{H.K.~Park}
\author[cau]{K.~Siyeon}
\author[ibs]{J.H.~So}
\author[inr]{V.I.~Tretyak}
\author[ibs]{Y.S.~Yoon\corref{cor1}}
\ead{ysy@ibs.re.kr}

\address[knu]{Department of Physics, Kyungpook National University, DaeGu 41566, Korea}
\address[cau]{Department of Physics, Chung-Ang University, Seoul 06974, Korea}
\address[ibs]{Center for Underground Physics, Institute for Basic Science,  Daejeon 34047, Korea}
\address[kriss]{Korea Research Institute of Standards and Science,  Daejeon 34113, Korea}
\address[inr]{Institute for Nuclear Research, MSP 03680 Kyiv, Ukraine }

\begin{abstract}
The first phase of the Advanced Mo-based Rare Process Experiment (AMoRE-I),
an experimental search for neutrinoless double beta decay (\znbb) of \Mo~in calcium molybdate (\CAMOO) crystal 
using cryogenic detection techniques, is in preparation at the YangYang underground laboratory (Y2L) in Korea.
A GEANT4 based Monte Carlo simulation was performed for 
the first-phase AMoRE detector and shield configuration.
Background sources such as \U, \Th, \U[235], and \Pb[210]~
from inside the crystals, surrounding materials, outer shielding walls of the Y2L cavity were simulated.
The estimated background rate in the region of interest was estimated to be \lt 1.5 \tmEm{3}  counts/keV/kg/yr (\UNIT).
The effects of random coincidences between backgrounds and two-neutrino double beta decays of \Mo~ 
as a potential background source were estimated to be \lt 2.3 \tmEm{4} \UNIT.
\end{abstract}


\end{frontmatter}

\linenumbers

\section{Introduction}\label{sec:intro}
As of today, on the basis of results from a number of neutrino oscillation experiments, it is
known that neutrinos have mass. 
However, their absolute mass scale is still not known \cite{Mohapatra:2005wg,Beringer:1900zz,Giunti:2007ry}. 
The half-life of neutrinoless double beta decay (\znbb) of certain nuclei is related to the effective Majorana
neutrino mass, and  the investigation of neutrinoless double beta decays is the only practical
way to determine the absolute neutrino mass scale and the nature of the neutrino such as Majorana or Dirac particle \cite{Giunti:2007ry}.
The Advanced Mo-based Rare Process Experiment (AMoRE) \cite{Bhang:2012gn} is an experimental search for
neutrinoless double beta decays of \Mo~nuclei using \CAMOO~(CMO) scintillating crystals operating
at milli-Kelvin temperatures and being planned to operate at the YangYang underground laboratory (Y2L) in Korea. 
The AMoRE experiment will run in a series of phases \cite{Alenkov:2015dic}; 
the first phase of the experiment (AMoRE-I) will use a \sm5 kg (possibly, up to 10 kg) array of CMO crystals. 
The goal of background level for AMoRE-I is 0.002 counts/keV/kg/yr (\UNIT) in the region of interest (ROI), 3.034 $\pm$ 0.01 MeV.
Radiations originating from the CMO crystals themselves are known to be the dominant source of backgrounds \cite{Artusa:2014wnl}.
Other possible sources are activities from radioisotopes contained in the \U, \Th, and \U[235]~decay chains from materials in the
nearby detector system 
and the outer lead shielding box that produce signals in the crystals. 
Backgrounds from more remote external sources such as the surrounding rock material and cement floor are also considered as possible sources.
Random coincidences of radiations from different background sources with two-neutrino double beta (\twonubb) decays of \Mo~in the CMO is known as a possible background source in the ROI \cite{Beeman:2011bg,Chernyak:2014ska}.
In this paper we estimate background rates due to the above mentioned background sources 
by performing simulations.

\section{The AMoRE-I Experiment}

\subsection{The geometry for detector simulation}
\begin{figure} [!htb]
\begin{center}
\begin{tabular}{ccc}
\includegraphics[width=0.3\textwidth, trim=10 50 10 20,clip ]{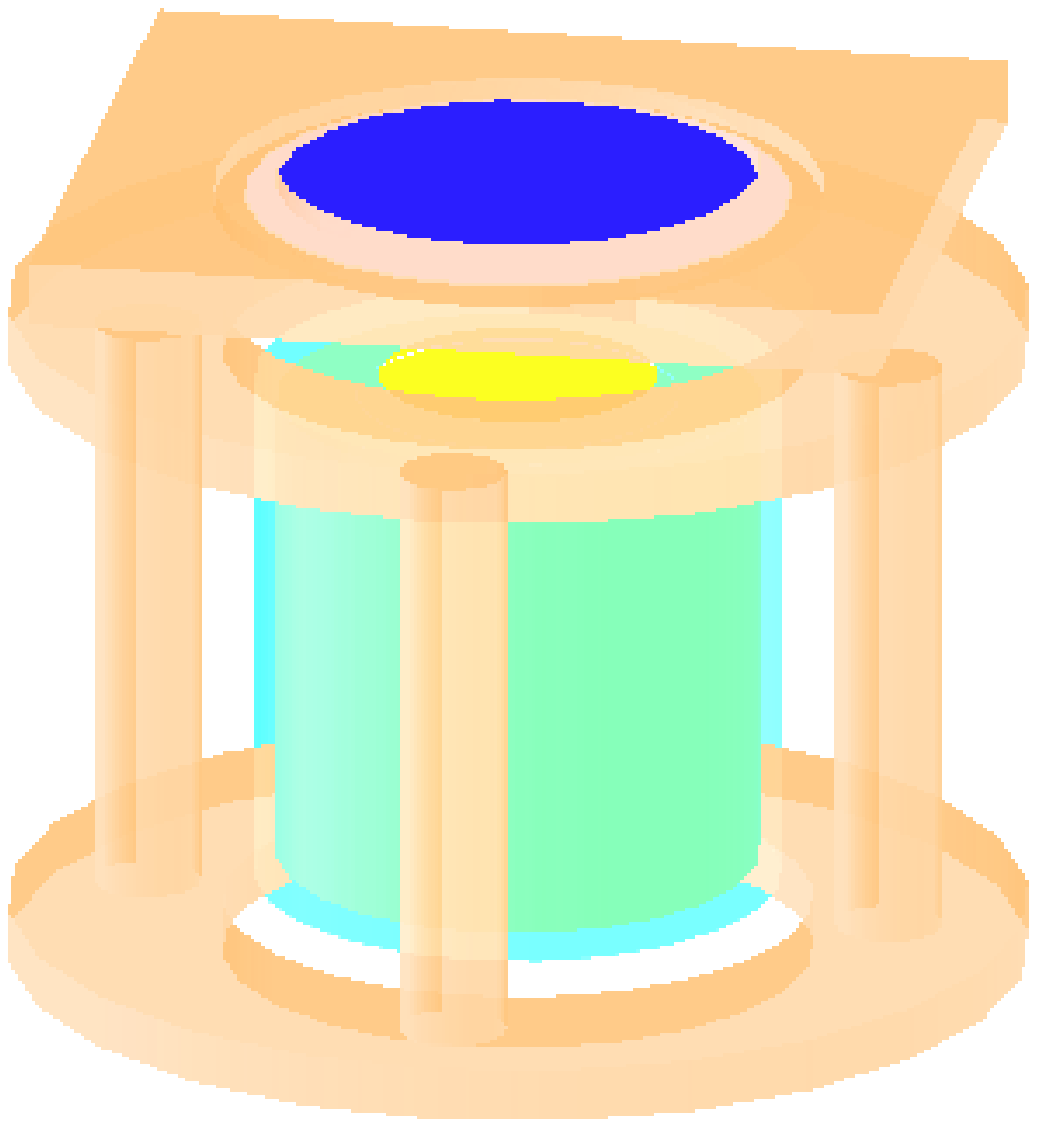} &
\includegraphics[width=0.27\textwidth, trim=10 10 10 20,clip ]{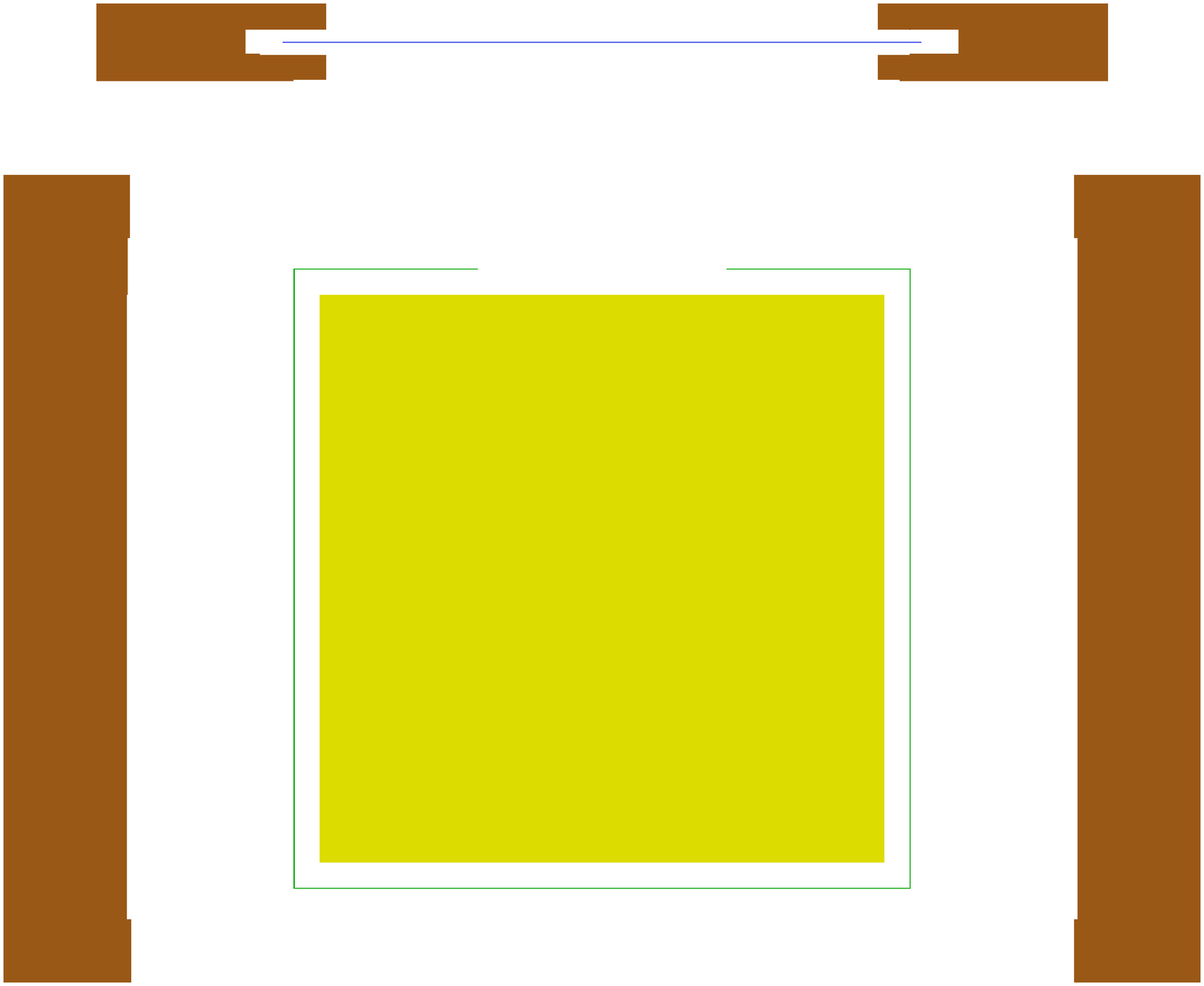} &
\includegraphics[width=0.3\textwidth, trim=10 20 10 0,clip]{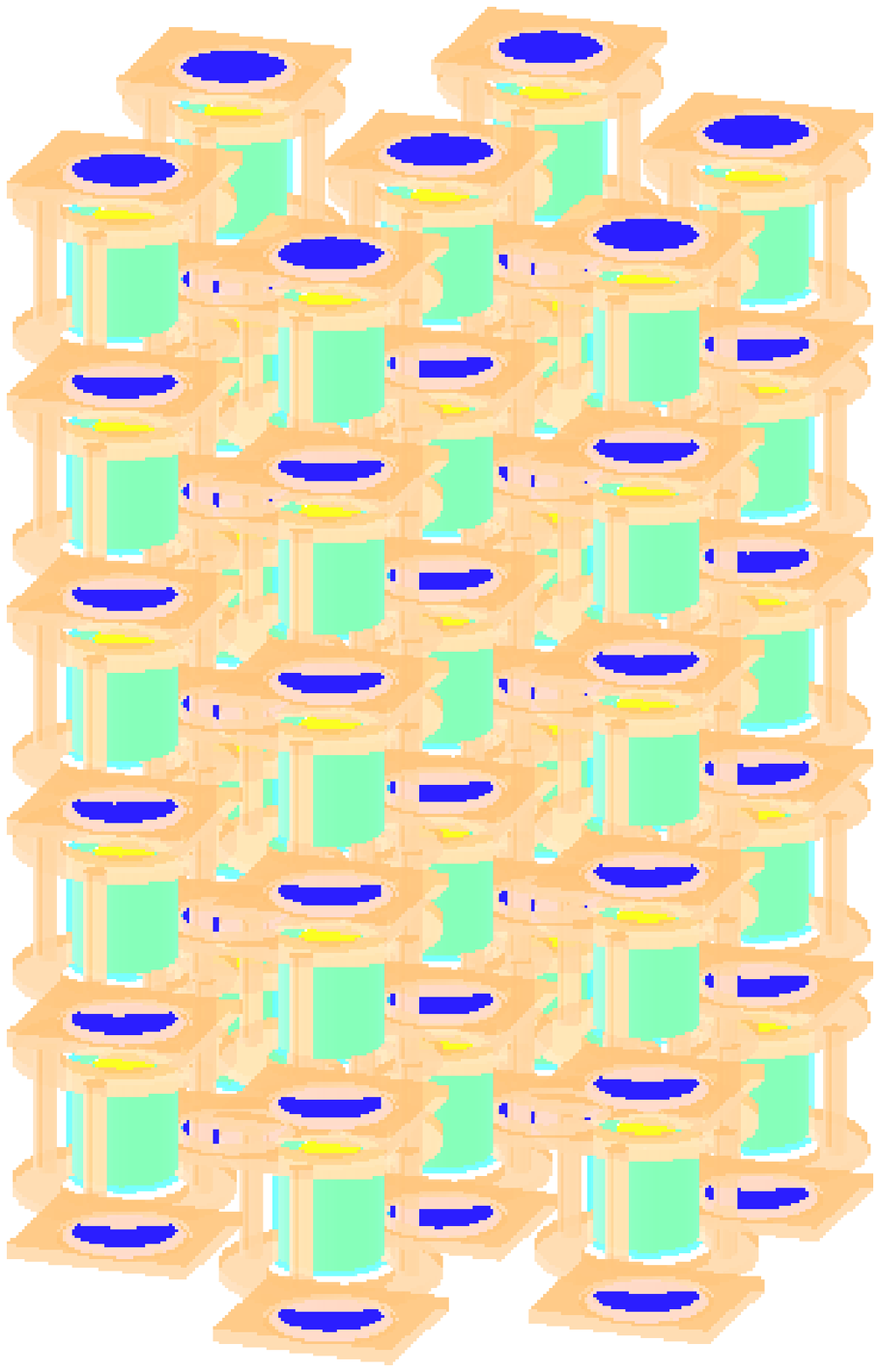} \\
(a) & (b) & (c)  \\
\end{tabular}
\caption[Crystals and Cu supporting frame]{
(a) A CMO crystal (yellow) with Vikuiti reflector (light green) and CMO supporting copper frame (brown).
Top of the reflector is open for a Ge wafer.
(b) Vikuiti reflector surrounding each crystal except for the center area on the top.
(c) 35 CMO crystals are stacked in 5 layers and 7 columns.
}
\label{fig:cmo}
\end{center}
\end{figure}

The detector geometry used for the AMoRE-I simulation includes thrity-five CMO crystals, 
shielding layers internal to the cryostat, an external lead shielding box, and outer rock walls of the Y2L cavity.
Each crystal has a cylindrical shape with 4.5 cm diameter, 4.5 cm height, and mass of 310 g.
The total mass of  crystals is 10.9 kg, originated from initial design of AMoRE-I experiment.
The thirty-five crystals are arranged in seven columns, each with five crystals stacked coaxially, 
with the center column surrounded by six external ones. 
The side, bottom and portion of top surfaces of each crystal are covered by 
a 65 $\mu$m-thick Vikuiti Enhanced Specular Reflector film (former VM2000) \cite{esr}.
Each crystal is mounted in a copper frame, which is a complicated design in reality, 
but, in the simulation, a simplified CMO supporting copper frame with corresponding mass is used, as shown in Fig. \ref{fig:cmo}. 
A Ge wafer and its copper frame are located above each crystal and below the lowest crystal.
More detailed detector description can be found elsewhere \cite{Alenkov:2015dic, Lee:2015tsa}.

\begin{figure} [!htb]
\begin{center}
\begin{tabular}{cc}
\includegraphics[width=0.44\textwidth, trim=160 30 210 150,clip ]{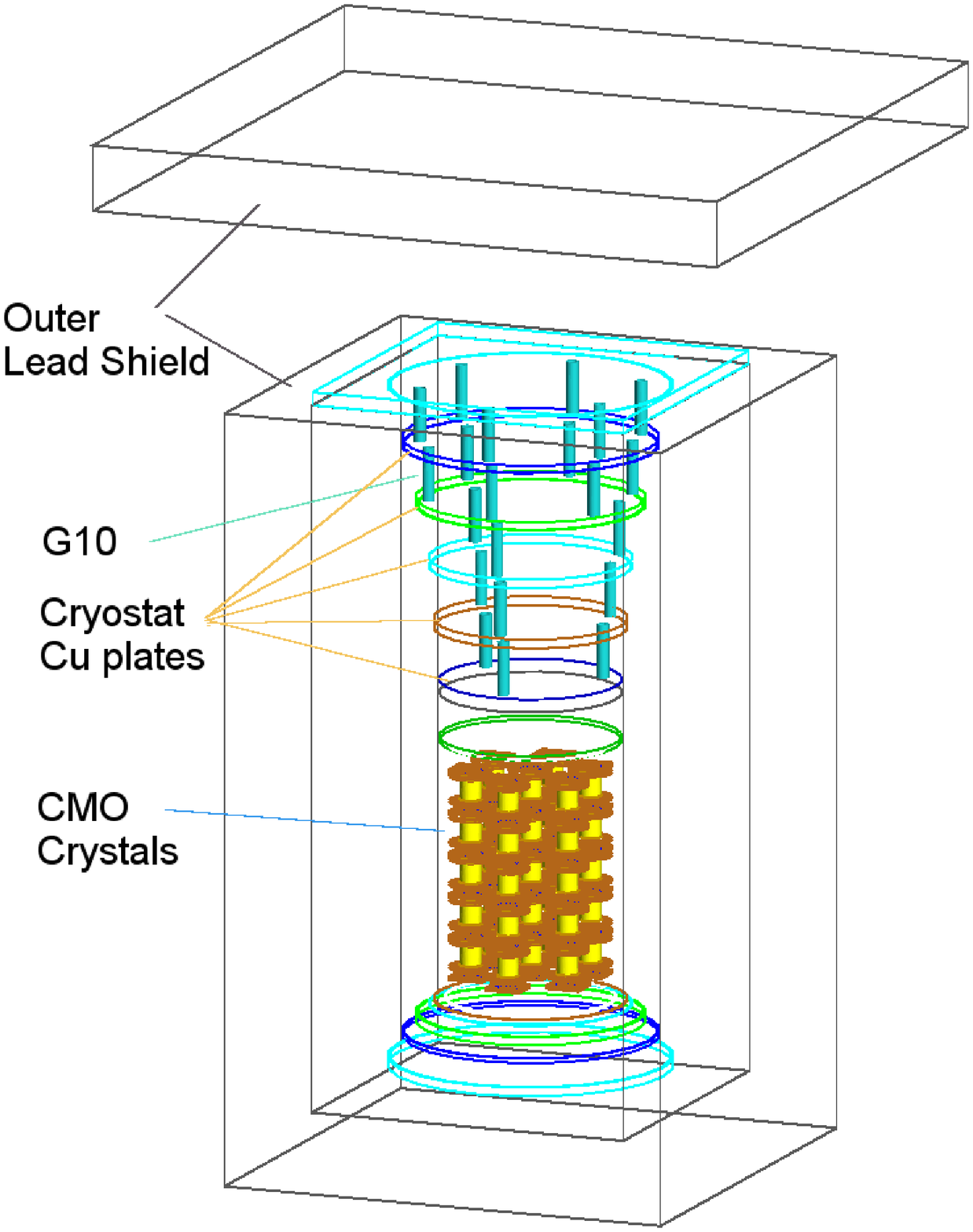} &
\includegraphics[width=0.44\textwidth, trim=180 80 300 270,clip ]{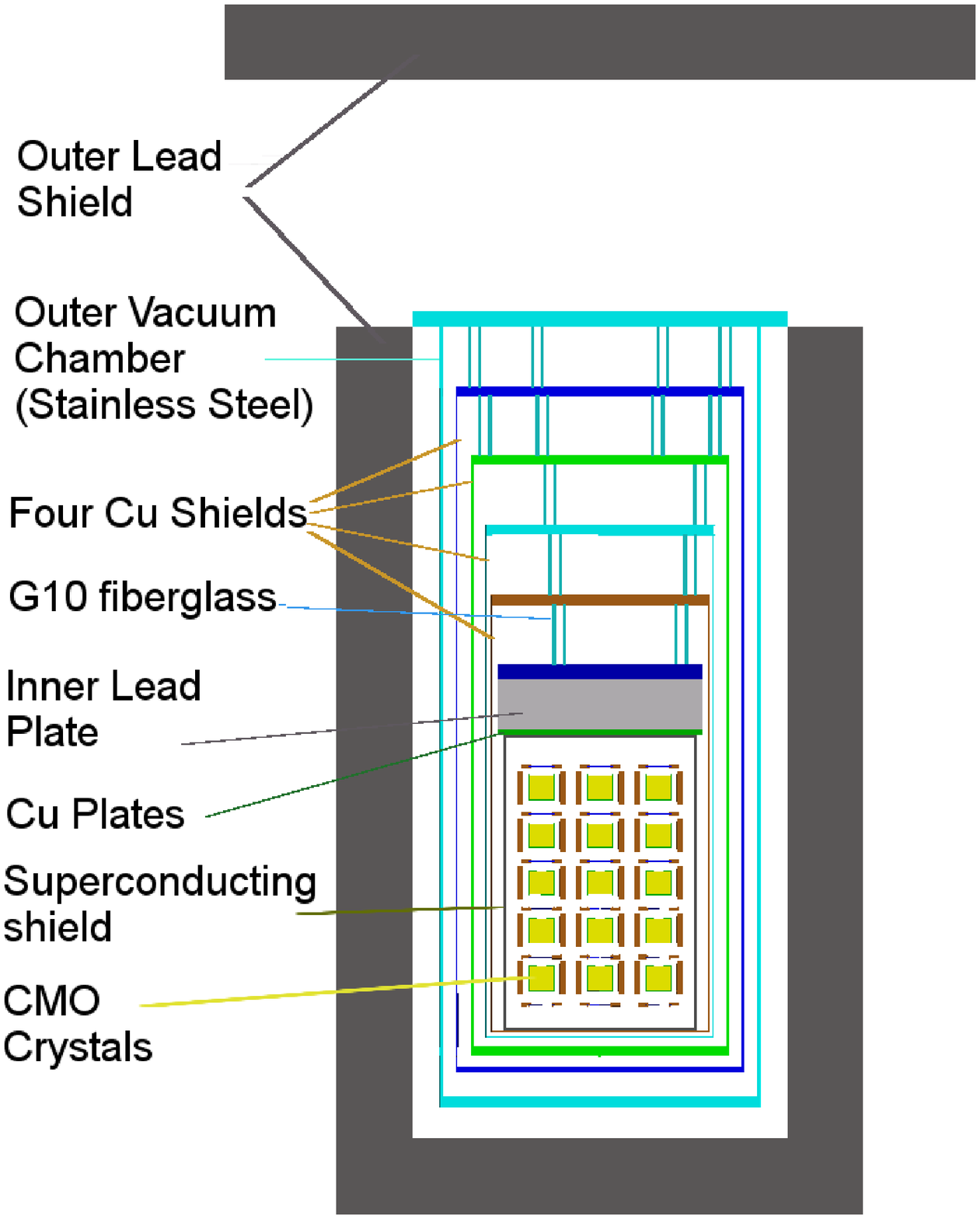} \\
(a) & (b)
\end{tabular}
\caption[Detector Geometry]{
Outside the cryostat, a 15 cm-thick outer lead shield is located.
The cryostat is composed of a stainless steel shield and four Cu shields.
Top plates of cryostat shields are connected with G10 fiberglass tubes (a). 
All the way inside the cryostat, an inner lead plate (gray) on a top of a Cu plate is located above crystals. 
A 2-mm-thick superconducting lead shield surrounds crystals inside the cryostat.}
\label{fig:shields}
\end{center}
\end{figure}

The whole crystal assembly is enclosed in a cylindrical 2-mm-thick lead superconducting magnetic shielding tube 
(diameter of 40.0 cm and height of 44 cm)
with top and bottom discs, which is made of an ultra-low activity, ancient lead, as shown in Fig. \ref{fig:shields}.
A 1-cm-thick copper plate and 10-cm-thick lead plate (diameter of 40.8 cm and mass of 148.3 kg)
are located in series above the CMO crystal assembly, 
to attenuate backgrounds from materials above crystals inside the cryostat, 
such as wires, temperature sensors, heaters, G10 fiberglass tubes, and stainless steel tubes.
The crystal assembly is placed inside of four concentric copper cylinders with total Cu thickness of 10 mm,
all within a 5 mm-thick outer stainless-steel vacuum cylinder, as shown in Fig. \ref{fig:shields}. 
Top plates of shields are connected with G10 fiberglass tubes,
which are made of a woven fiberglass material with 12 cm-height, 2.5 cm-diameter (\sm 25 g)
and are known as a material with high radioactive background.
In this simulation, simple top plates and G10 fiberglass tubes are taken into account in the detector geometry.
Realistic structures and features will be included on top plates in future simulations.

\begin{figure} [!htb]
\begin{center}
\includegraphics[width=0.45\textwidth, trim=40 130 30 15,clip ]{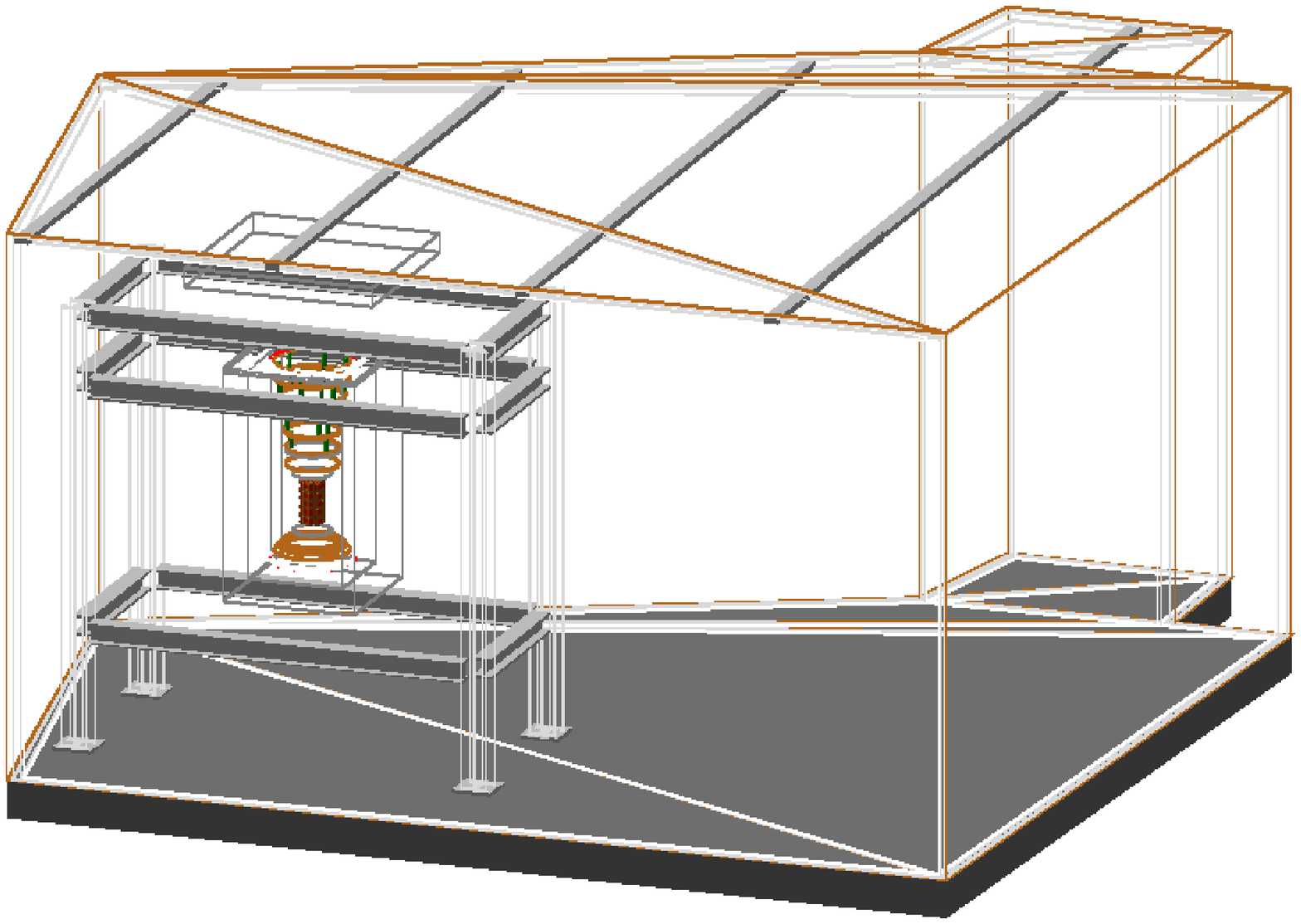}
\caption[Laboratory Geometry]{
 Cement floor, laboratory walls, ceiling of Y2L laboratory. 
 The ceiling and walls are covered by plaster board and steel plate in series.}
\label{fig:lab}
\end{center}
\end{figure}

The cryostat is located inside a 15-cm-thick external lead shield, a total mass of \sm 15.6 ton.
A top lead shield plate, covering area of 150\tm150 \cmsq, is placed \sm50-cm above the top of the cryostat
gives a space for pipes and valves of the cryostat in the detector system. 
For radiations from the rock walls surrounding the experimental enclosure, 
the simulation uses a 50-cm-thick spherical rock shell, which is an optimized thickness for saturated escaping $\gamma$-rays. 
For radiations from the laboratory environment such as the cement floor, the laboratory walls and ceiling, 
and iron supporting system, simplified designs are implemented into the geometry as shown in Fig. \ref{fig:lab}.

\subsection{Simulation method}

We have performed simulations using the GEANT4 Toolkit \cite{Agostinelli:2002hh}. 
In internal or external materials, 
radioactive sources such as full decay chains of \U, \Th, and \U[235]~are simulated.
In general, most decay sources and their daughters are considered to be in equilibrium states,
thus all related activities within the chains are simply equal to \U, \Th, and \U[235]~activities 
multiplied by the branching ratios for decays of the daughter isotopes.
However, for \U~decays inside crystals, a broken decay chain is considered by 
measured \U~and \Rn[222]~concentrations separately.
For radiations from rock, instead of \K[40] and full decay chain simulation of \U~and \Th~nuclei, 
\g s with energy of 1.46, 1.87, and 2.61 MeV, are simulated at the rock shell, 
since daughter nuclei, \a, and \b~cannot penetrate the lead shield and the cryostat, made of stainless steel and copper shields.

Each simulated event includes deposits inside crystals within an event window of 100 ms when a decay occurs,
which is based on a few times of a typical pulse width (\sm20-30 ms) in cryogenic measurements \cite{Lee:2015tsa}.
Sometimes decays with relatively short half lifetime such as \Po[212]~decay with a half lifetime of 300 ns
and the followed decays will appear in the same event, called pileup events, 
and they were treated as one event in simulated event.
Furthermore, effect of random coincidence of \Mo~\twonubb~decay events and other radioactive sources inside crystals
were estimated by a convolution technique. 
For random coincidence rate calculation, \Mo~\twonubb~decay events were generated 
using an event generator, DECAY0 program \cite{Ponkratenko:2000um}, 
and those events were used as input particles in our simulation package with the AMoRE-I detector configuration 
to get a distribution of \Mo~\twonubb~decays inside the CMO crystals.

\section{Analysis}
Radioactive sources are simulated in the following materials which are known as dominant background sources:
(i) First, the internal background in CMO crystals
(ii) Second, backgrounds from materials in the detector system, 
including CMO supporting copper frame, Vikuiti reflector, superconducting lead shield, Cu plate under internal lead,
internal lead plate, G10 fiberglass tubes, and outer lead shielding box
(iii) Third, backgrounds from rock material and surrounding underground laboratory.
Then, we estimate an anti-coincidence rate and a random coincidence rate with the \Mo~\twonubb~decays.

Anti-coincidence rate is estimated from events in the ROI in one crystal, called a single hit event,
while events with hits in more than one crystal are rejected because they are obviously from radioactive decays.
The single hit events are surface \a~events, \b-\a~pileup events, and \b-like (\b~or \g)~events.
First, the surface \a-events from crystals or near-by materials can deposit energies in the crystals
in a continuum distribution up to Q-value (see Section \ref{nearby} and Ref. \cite{Alessandria:2012zp}),
which can appear in the ROI.
Those \a~signals are distinguished from \b~and \g~signals by a pulse shape discrimination (PSD) 
and a separation power of 7.6 \sgm~between \a~and \b/\g~events were reported \cite{Kim:2015pua}.
We assume that the \a~event rejection power is 99.9999 \% in this estimation.
Second, the \b-\a~pileup events are from decays of (\Bi[212]$+$\Po[212]) in \Th~decay chain, with half lifetime of 300 ns,
and decays of (\Bi[214]$+$\Po[214]) in \U~decay chain, with half lifetime of 164 $\mu$s.
When the \b-\a~pileup events occur near the surface of crystals or near-by materials, 
events with a faction of \a~Q-value and \b-energy can appear in the ROI.
These \b-\a~pileup events from \U~and \Th~chains can be rejected by the PSD analysis with a rejection power of 99 \%,
which was reported with a clear separation between \a~and \b-like events with a prototype detector \cite{Kim:2015pua, Alenkov:2015dic}.
In this estimation, 90 \% rejection efficiency (\sm1.6 \sgm) is assumed, conservatively.
After the surface \a~rejection, remaining anti-coincidence rate is mainly \b-like single hit event rate in the ROI.
A further rejection scheme for \b-like single hit events will be discussed in Section \ref{internal}.

Random coincidence events are originated from any two signals, 
while the \b-\a~pileup events are from two consecutive decays with relatively short half lifetimes.
When two random signals occurred within a time resolution, they might be considered as one pulse, 
which could be a background source in the ROI.
Random coincidence event rate within a time resolution of 0.5 ms
were estimated by a convolution of two background spectra \cite{gbkimthesis}.
The Random coincidence rate includes not only randomly coincident events by two \Mo~\twonubb~decays,
but those by a \Mo~\twonubb~decay and another decays from other background sources (see Section \ref{rcrate}).

Activities of \U, \Th, and \U[235]~which are used to normalize the simulation results,
were measured by a germanium counting and inductively coupled plasma mass-spectroscopy (ICP-MS). 
The High Purity Germanium (HPGe) measurements were performed at the Y2L \cite{2016JPhCS.718f2050S}. 
The ICP-MS measurements were all performed by the KAIST Analysis Center for Research Advancement (KARA), Korea.
The activity of backgrounds in a CMO crystal was measured by low temperature detector technique \cite{Kim:2015pua}.
Activities of some materials were from published references.
The concentration of materials are listed in the following section.
Here, the activities from bulk and surface contaminations are considered as uniform for crystals and near-by materials.
The surface alpha events are also from external contaminations \cite{Alessandria:2011vj}, which will be studied in the future.

\subsection{Internal background in CMO crystals}

\subsubsection{Background rate due to sources inside CMO crystals}
\label{internal}
\begin{figure}[!htb]
\begin{tabular}{cc}
\includegraphics[width=0.48\textwidth]{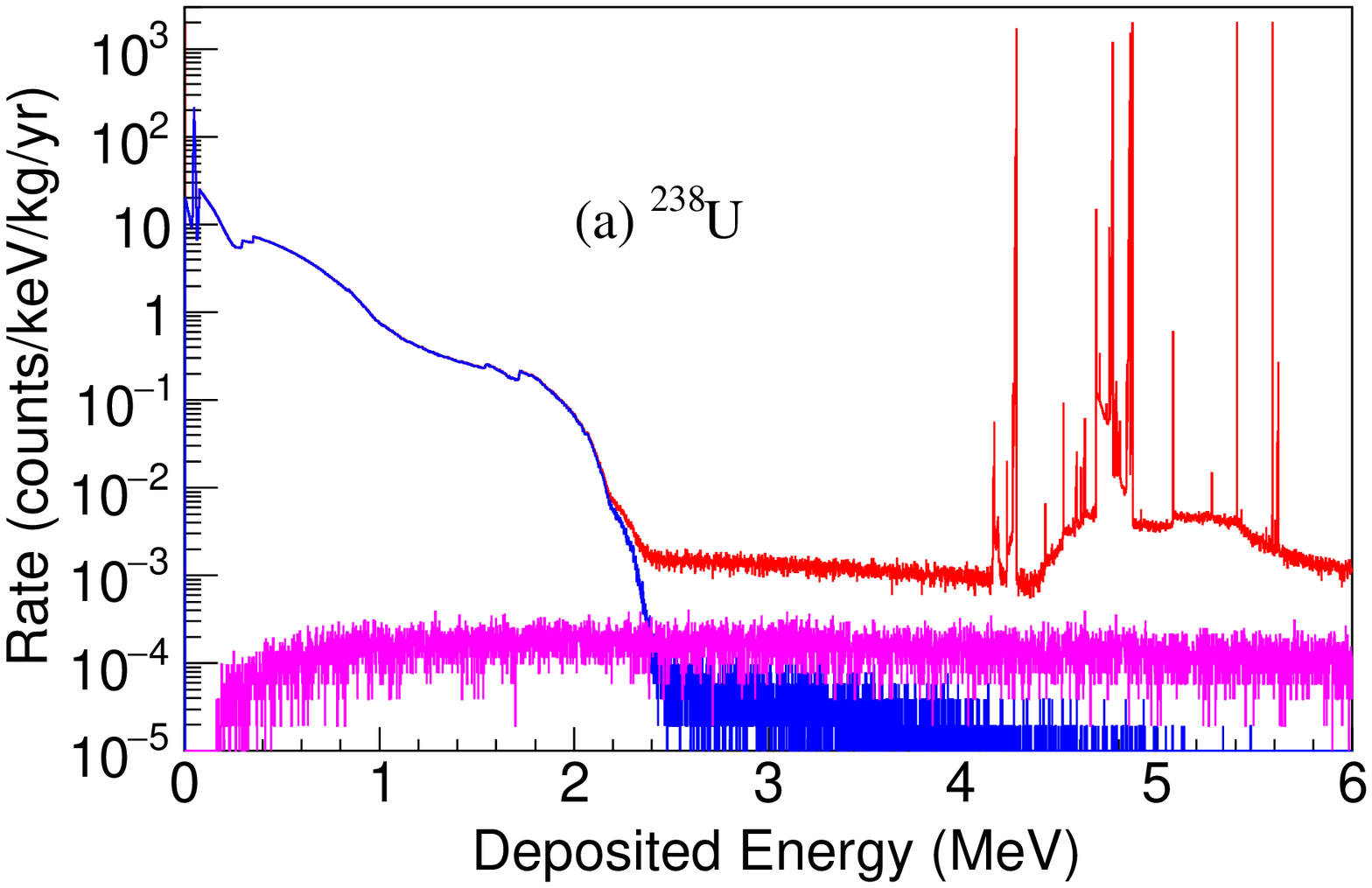}
\includegraphics[width=0.48\textwidth]{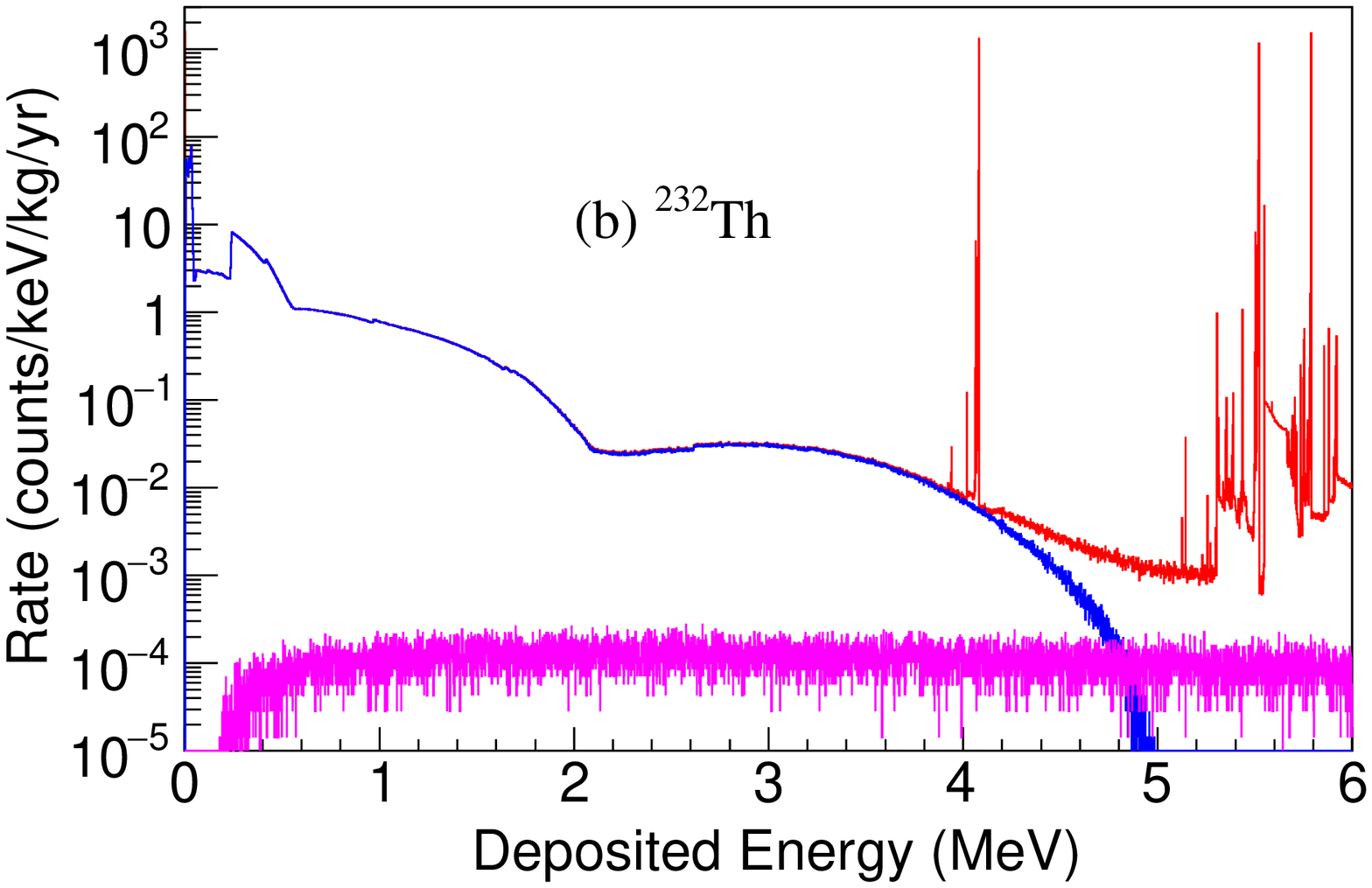}
\end{tabular}
\caption[CMO]{
Single hit event rate distribution from (a)\U~and (b)\Th~inside crystals for
total events(red) including \a s, \b-\a~pileup events (pink), and \b-like event exclusively (blue).
The difference between the total rate (red) and sum of the others is \a~particle event rate.
}
\label{ref:bg_accu}
\end{figure}

The activities inside crystals, listed in Table \ref{tab_conc}, are 
from a recent measurement of a CMO crystal \cite{Kim:2015prep}, except the activity of \Th.  
For the activity of \Th, a conservative upper limit are used. 
Using activities of radioactive sources in crystals, event rates in the ROI are normalized to
the unit of counts/keV/kg/yr (ckky).
Fig. \ref{ref:bg_accu} shows deposited energy distributions in CMO crystals for total single-hit event rate, 
\b-\a~pileup rate, and \b-like event rate exclusively.

\begin{table} [!htb]
\begin{center}
\caption{Activities based on a CMO crystal measurement \cite{Kim:2015prep}  [mBq/kg]}
\label{tab_conc}
\begin{tabular}{c|c|c|c|c|c}
    \hline
     &	\Pb[210]   & \U & \Ra(\Rn[222])   & \Th  & \U[235](\Bi[211]) \\ 
    \hline \hline
     Activity  & 7.3 & 0.98 & 0.065 & \lt0.05 & 0.47 \\ 
    \hline
  \end{tabular}
\end{center}
\end{table}

As shown in Fig. \ref{ref:bg_accu}, \b-like events from \Tl~decay in \Th~chain 
is a dominant source candidate in the ROI, but they can be rejected
using a time correlation with the \a~signal from the preceding \Bi[212]$\rightarrow$\Tl[208]~\a~decay, called \a-tagging method.
Rejection of all events occurring within 30 mins after an \a~event with 6.207 MeV in the same crystal,
results in a 97.4\% veto efficiency for \Tl-induced \b~events in the \Mo~\twonubb~signal region,
while introducing \sm 1\% dead-time.

\subsubsection{Random coincidence rate of two \Mo~\twonubb~decays}
\label{rcrate}
The \twonubb~decay in a CMO approaches zero rate at the end-point energy, 
but random coincidence of these events can sum together (pileup) creating backgrounds for the \znbb~signal.
The random coincidence of two \Mo~\twonubb~decays in an energy range, $\Delta E$, 
can be expressed as \cite{Chernyak:2012zz},
\begin{equation}
R_{rc} = \tau \cdot R_{0}^{2} \cdot \varepsilon 
 = \tau ~\Big( \frac{\rm{ln} 2~ N}{T^{2\nu2\beta}_{1/2} } \Big)^{2} ~\varepsilon,
\end{equation}
where $\tau$ is a time resolution, $R_{0}$ is a decay rate, N is the number of \Mo~decays and $\varepsilon$ is the probability of appearing events in the
$\Delta E$ interval, which was evaluated by convolution of two simulated distributions here.
The expected rate of \twonubb~decay in a single CMO crystal with \sm 306 g is 0.0028 counts/s, 
which is one double beta decay event per \sm6 mins.

\begin{figure}[!htb]
\begin{center}
\includegraphics[width=0.48\textwidth]{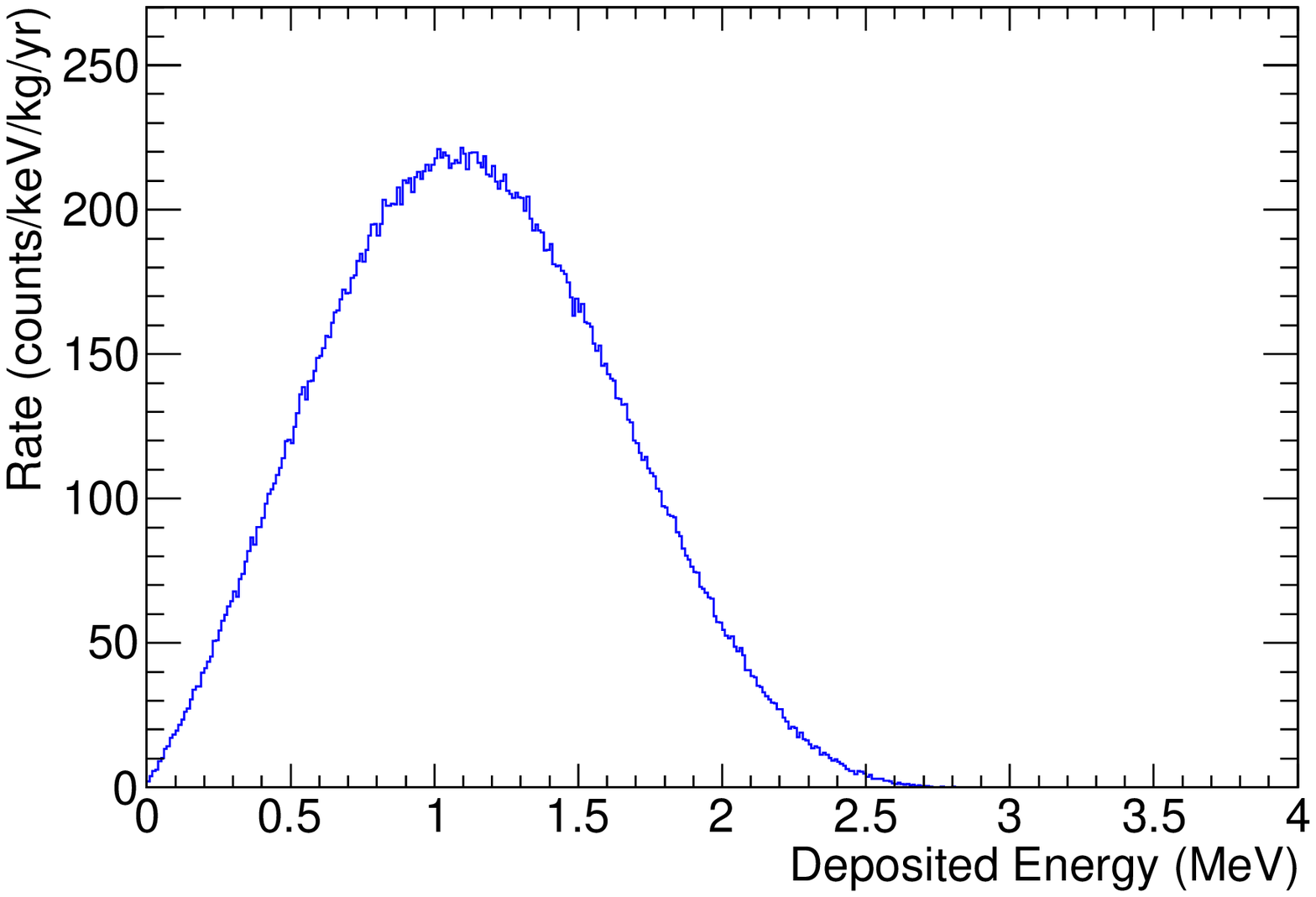}
\includegraphics[width=0.48\textwidth]{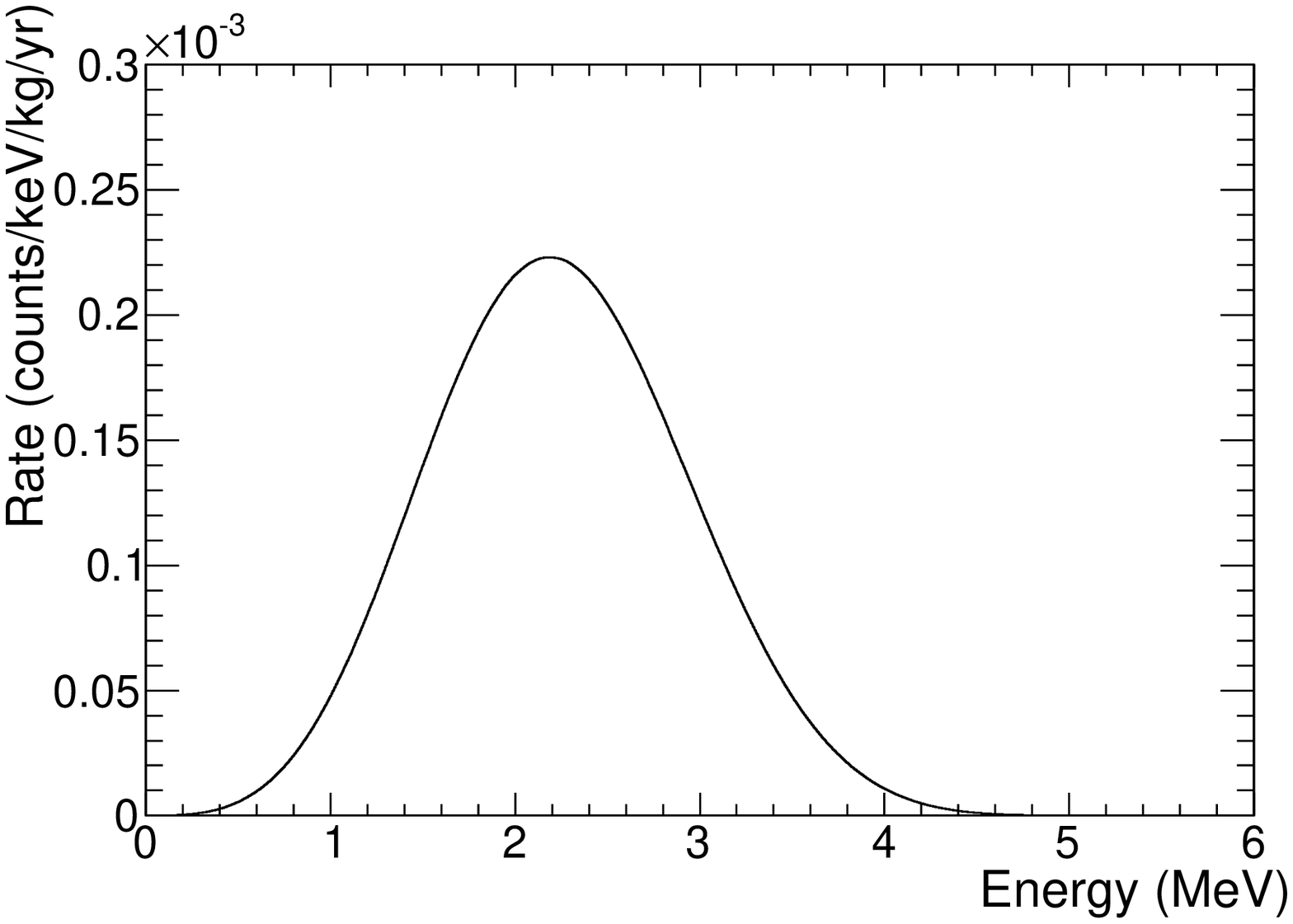}
\caption[ConvolutionBB]{The energy distribution  of \Mo~\twonubb~decays (left) and 
random coincidences of two \twonubb~decays of \Mo~(right). 
The random coincidence spectrum is derived by convolution of the \twonubb~spectrum, 
and normalized according to the \twonubb~rate assuming an 0.5 ms coincidence window.}
\label{fig:twobb}
\end{center}
\end{figure}

When 5\tmE{8} \twonubb~events were simulated in the CMO crystals, 
\sm 99 \% of events have energy deposits in one crystal (single-hit events) 
and the remaining events ($<$1 \%) produced hits in multiple crystals. 
The random coincidence rate of two \twonubb~decays was calculated by convolving two single-hit
\twonubb~decay energy distributions  (see Fig.~\ref{fig:twobb})
and the accidental rate in the coincidence window is 1.2 \tmEm{4} \UNIT~in the ROI.

Random coincidence rates of \Mo~\twonubb~decay and each radioactive background source 
inside CMO crystals such as \Pb[210], \U, \Th, \K, and \U[235]~are calculated as well.
Most concentration values are from the measurements shown in the Table \ref{tab_conc}, 
and \K~concentration is assumed as 1 mBq/kg \cite{Belli:2010zzc,Blum:1992vr}. 
The sum of random coincidence rates from Mo~\twonubb~decays and other nuclei is 7.80 \tmEm{6} \UNIT. 
In the same way, the random coincidence rates of two radioactive background sources inside CMO crystals
are calculated and the sum of rates between two background sources are 2.23 \tmEm{8} \UNIT.

\subsection{Backgrounds from materials in detector system}

\subsubsection{Backgrounds from materials inside and on the cryostat}
\label{nearby}
Activity of radioactive sources in Vikuiti reflector, measured by the HPGe at the Y2L \cite{Leonard:2016prep}, 
is  0.91 mBq/kg and 0.48 mBq/kg for \U~(\Ra[226]) and \Th~(\Th[228]), respectively.
The CMO supporting copper frame is made of a NOSV grade copper from Aurubis Co. and 
reported concentration of the NOSV copper \cite{Laubenstein:2009} is shown in Table \ref{bkg_vc}.
We purchased T2FA lead bricks from Lemer Pax with a certified activity of \Pb[210] 0.3 Bq/kg
for the inner lead plate and the superconducting lead shield. 
Activities of \U~and \Th~in the G10 fiberglass tubes are measured by the ICP-MS, as shown in Table \ref{bkglead}, 
which are in the similar order of the reported results \cite{SNO91_15}. 
The activities of \U~and \Th~in the outer vacuum chamber, made of stainless steel, are from a reference\cite{Artusa:2014wnl},
and a measurement of the stainless steel activity is being planned.

\begin{table} [!htb]
\begin{center}
\caption{Activities of \U~and \Th~in Vikuiti reflector  \cite{Leonard:2016prep} and NOSV copper  \cite{Laubenstein:2009}}
\label{bkg_vc}
\begin{tabular}{c|c|c}\hline
  & \U~(\Ra[226]) & \Th~(\Th[228]) \\ \hline \hline
Vikuiti Reflector & \lt 0.91 mBq/kg & \lt 0.48 mBq/kg  \\ \hline
CMO supporting &  \multirow{2}{*}{\lt16 $\mu$Bq/kg} & \multirow{2}{*}{\lt25 $\mu$Bq/kg}  \\
copper frame	&	&    \\ \hline
\end{tabular}
\end{center}
\end{table}

\begin{table}[!htb]
\begin{center}
\caption{Concentrations and activities of radioactive sources in materials inside and on the cryostat}
\label{bkglead}
\begin{tabular}{c|c|c|c}\hline
  &	\Pb[210] & \U & \Th  \\ \hline \hline
Superconducting  & \multirow{2}{*}{0.3 Bq/kg} & \multirow{2}{*}{1 ppt} & \multirow{2}{*}{1 ppt} \\ 
lead shield  &	&	&   \\ \hline
Cu plate  & - &  \multirow{2}{*}{\lt16 $\mu$Bq/kg} &  \multirow{2}{*}{\lt25 $\mu$Bq/kg}   \\ 
under inner lead &	&	&   \\ \hline
Inner lead plate & 0.3 Bq/kg & 1 ppt & 1 ppt   \\ \hline
G10 fiberglass	tubes & - & 1,732 ppb & 12,380 ppb   \\ \hline
Stainless steel shield &  & \lt0.2 mBq/kg & \lt0.1 mBq/kg  \\ \hline
Outer lead shield & \lt59 Bq/kg & 6.9 ppt & 3.8 ppt  \\ \hline
\end{tabular}
\end{center}
\end{table}

\begin{figure}[!htb]
\begin{center}
\begin{tabular}{cc}
  \includegraphics[width=.48\linewidth]{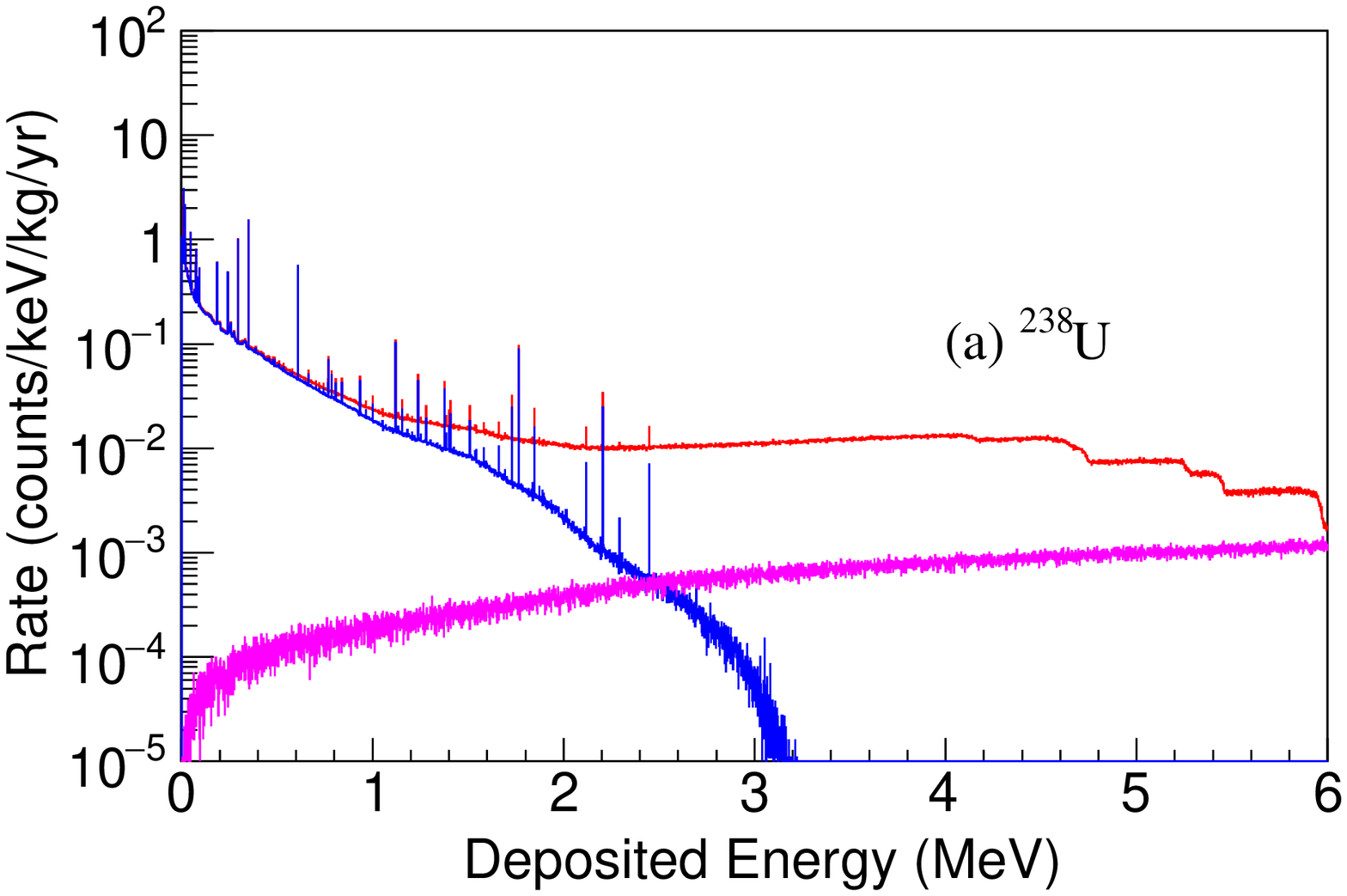} &  
  \includegraphics[width=.48\linewidth]{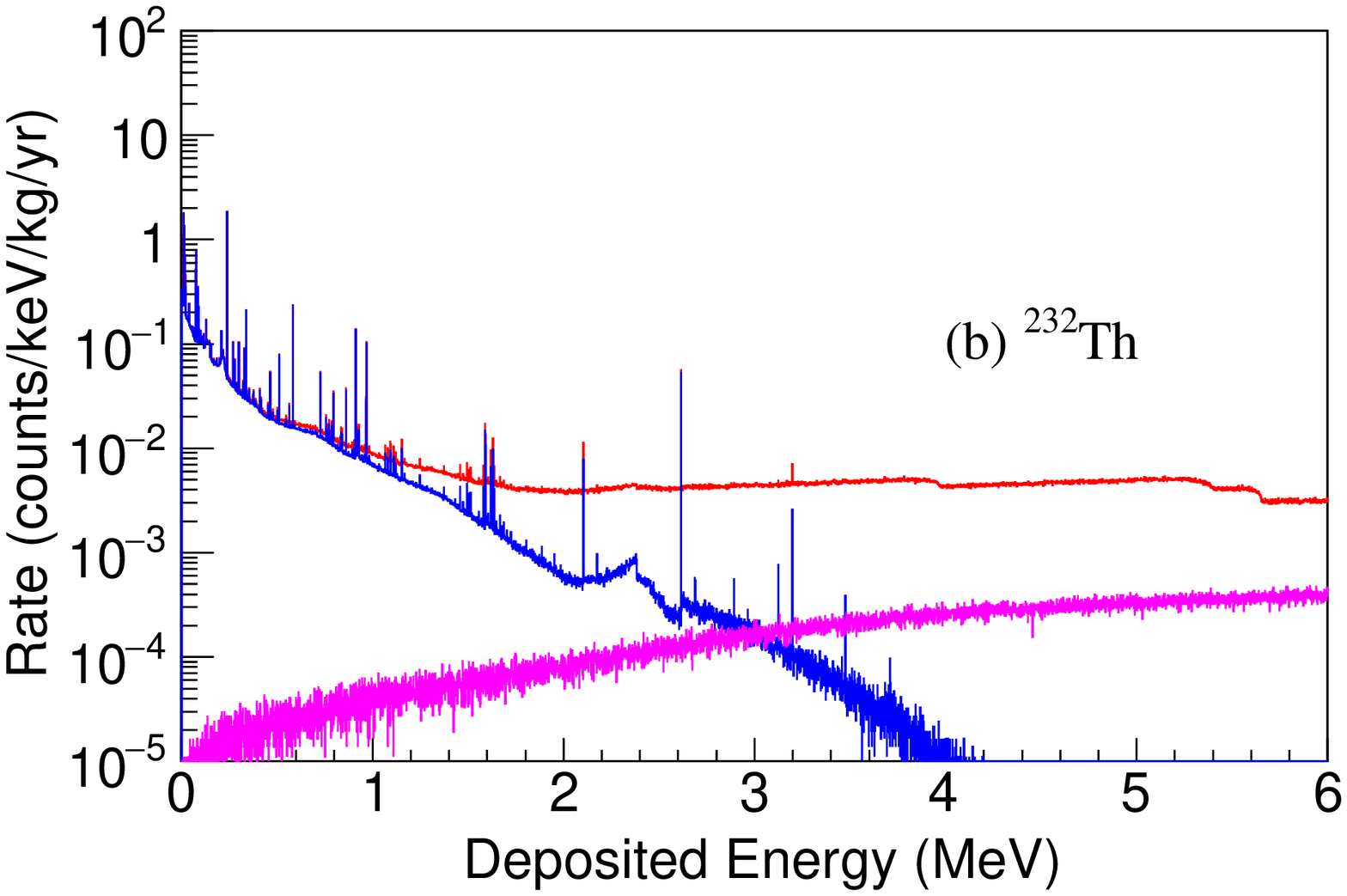} \\
  \includegraphics[width=.48\linewidth]{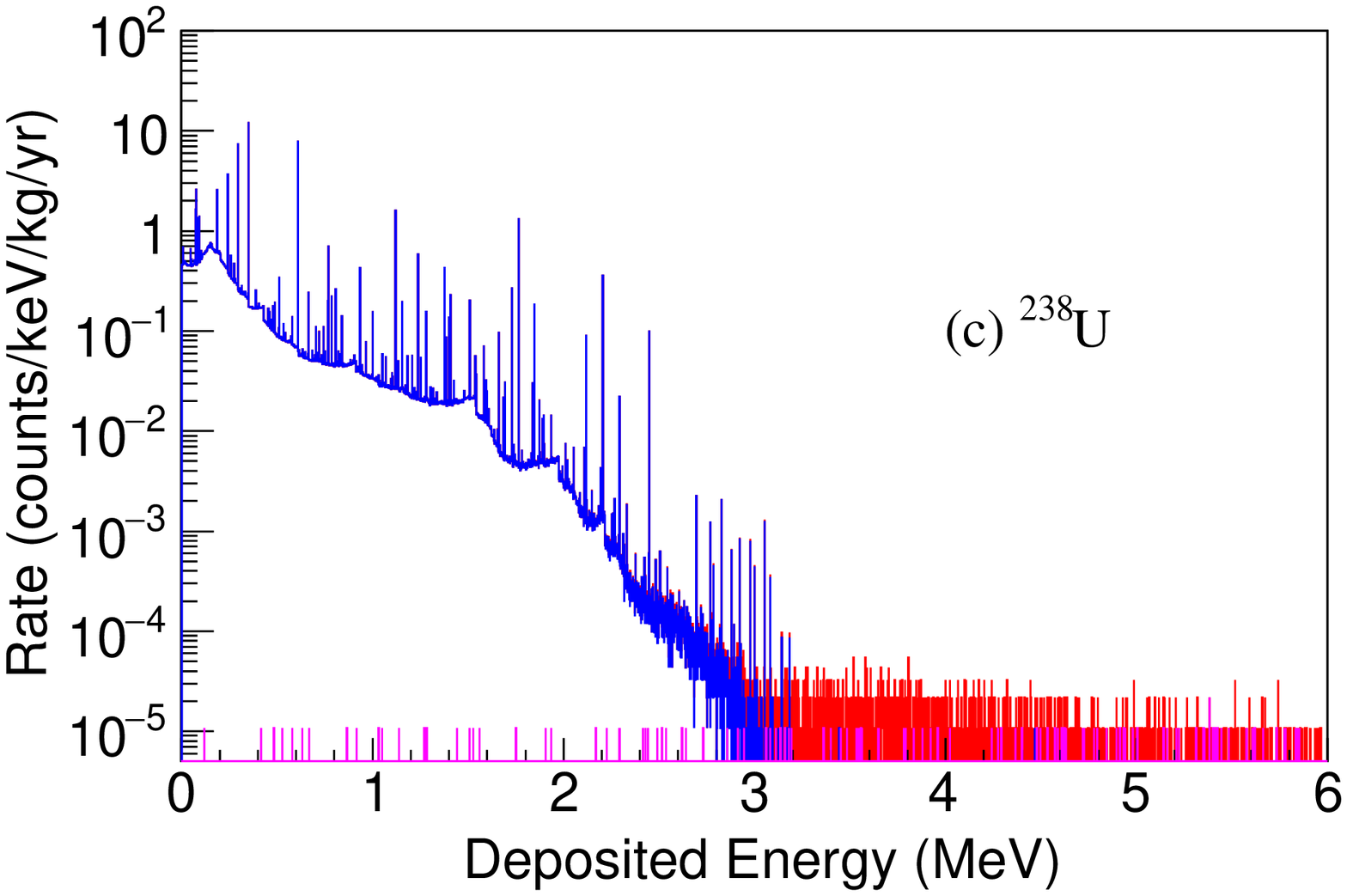} &  
  \includegraphics[width=.48\linewidth]{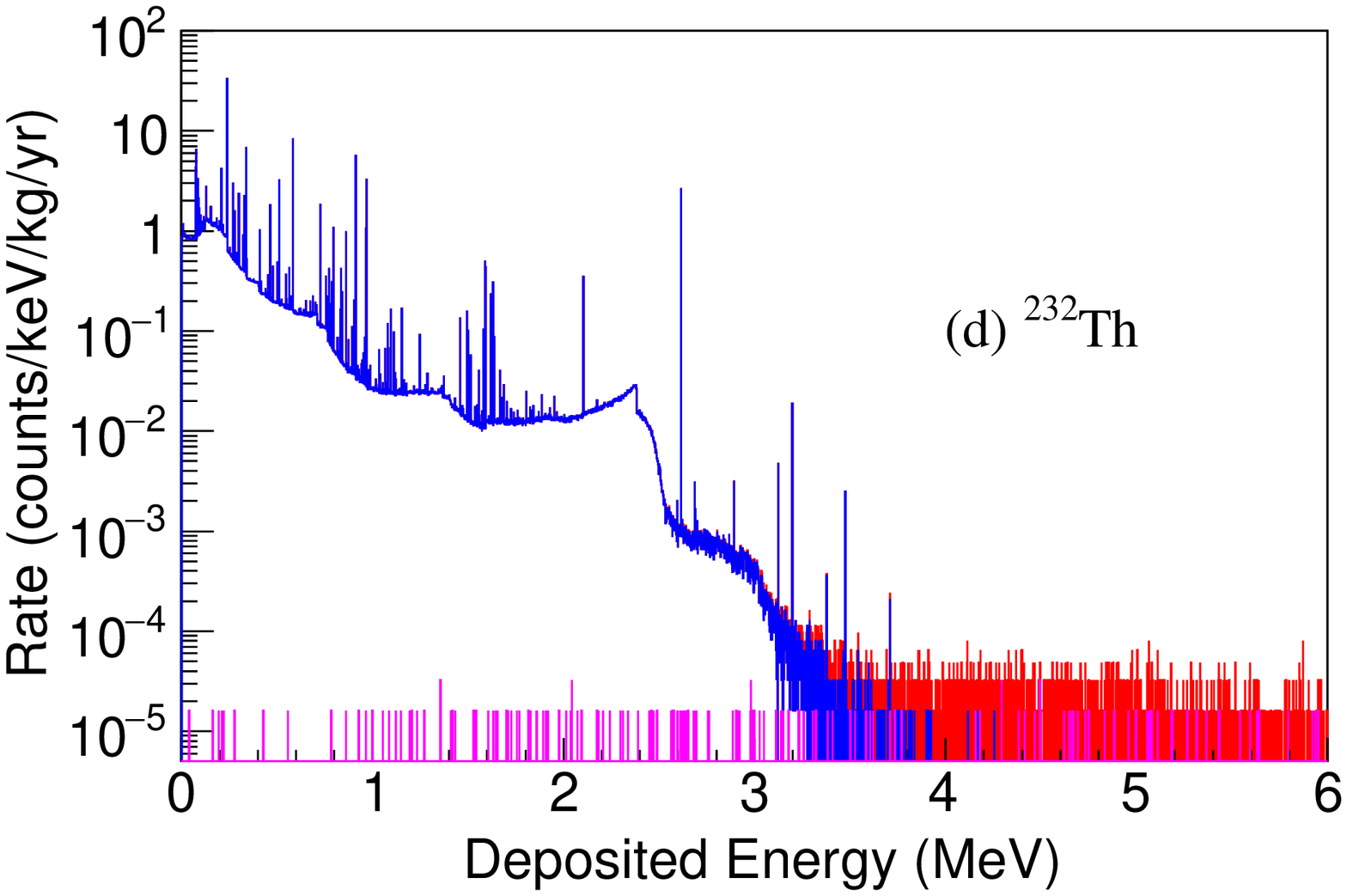} \\
\end{tabular}
\caption[CMOSupports]{Deposited energy of single hit events from \b~decay chain (blue) and 
all decay chains including surface \a~(red) in CMO crystals 
originated from \U~and \Th~in (a,b) Vikuiti reflector and (c,d) CMO supporting copper frame.}
\label{ref:bg_support}
\end{center}
\end{figure}

The estimated \a~and \b-\a~pileup event rates of \U~and \Th~contaminants in the Vikuiti reflector are in the order of \Em{4} \UNIT.
But they can be rejected by the PSD analysis like those internal background events.
The \b-like event rate is in the order of \Em{4} \UNIT~in the ROI.
The estimated \a~and \b-\a~pileup event rate in the CMO supporting copper frame are in the order of \Em{5} \UNIT.
Since each crystal is surrounded by the Vikuiti reflector except the top center area,
\a-particles and \b s from the CMO supporting copper frame rarely reach to crystals directly and
\b-like events, mostly \g-rays, make signals on crystals, as shown in Fig. \ref{ref:bg_support}. 
 
\begin{figure}[!htb]
\begin{center}
\begin{tabular}{cc}
\includegraphics[width=0.95\textwidth]{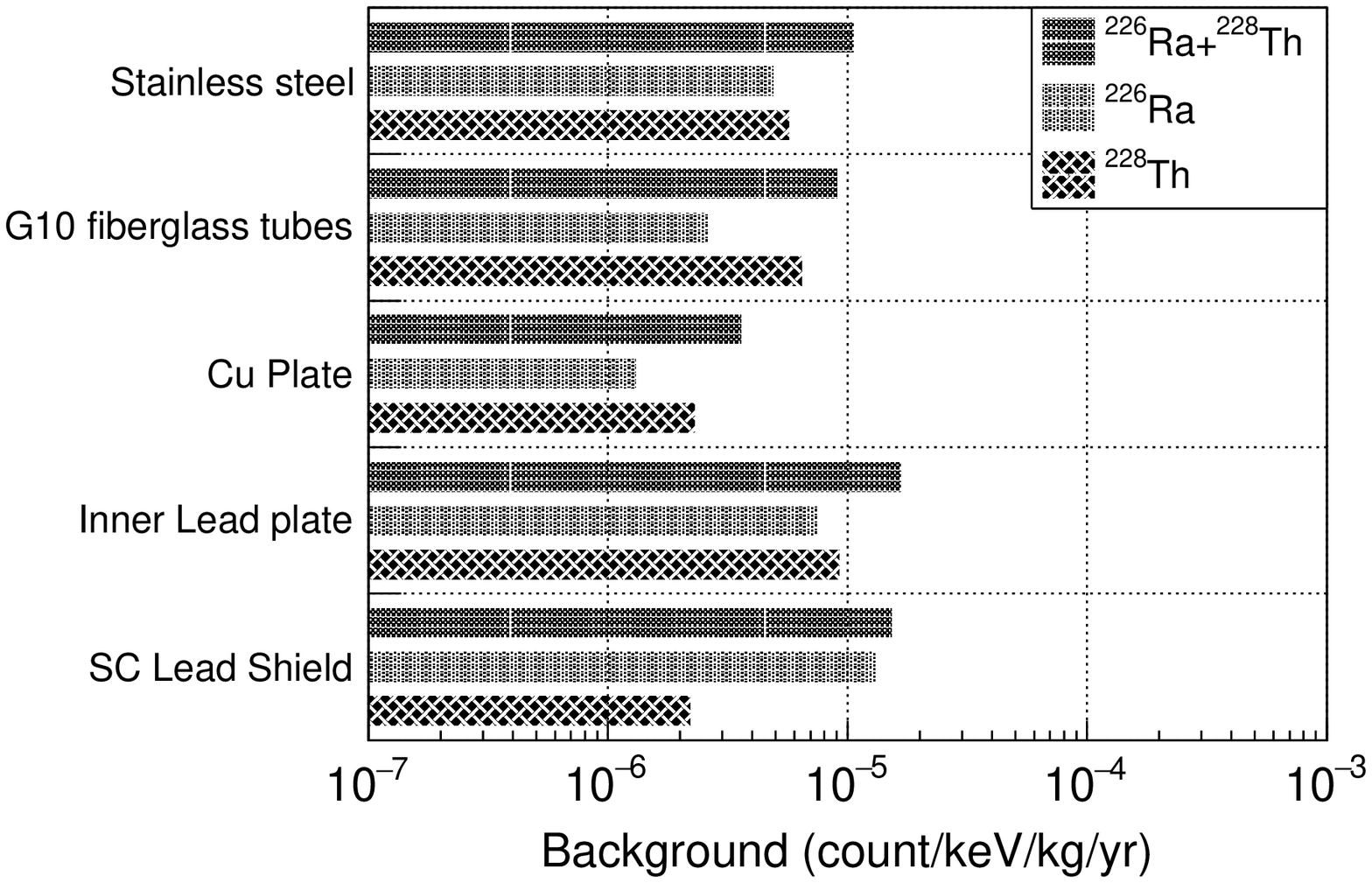}
\end{tabular}
\caption[External materials]{
Background \b-like event rates from \Ra~and \Th~in the ROI from the cryostat and shielding materials.
}
\label{ref:bg_external}
\end{center}
\end{figure}

The effect of radioactive sources from the superconducting (SC) lead shield and other materials, placed out side of the SC lead shield,
are from \b-like events, mainly \g s, like the CMO supporting copper frames.
When no events are found in the ROI for a few thousand years of exposure, 
an upper limit (90$\%$ C.L.) is estimated and 
the limit is to be in the order of \Em{5} \UNIT~by increasing statistics,
as shown in Fig. \ref{ref:bg_external}.

Random coincidence rates of \b-like events from materials inside the cryostat with \Mo~\twonubb~decay are 
in the order of  \sm\Em{7}$-$\Em{9} \UNIT~
for CMO supporting copper frame, SC lead shield, Cu plate, internal lead plate,
which are less than the random coincidence rate of two \Mo~\twonubb~decays inside the crystals. 

\subsubsection{Backgrounds from lead shielding box outside the cryostat}
\label{leadshield}
A major concern of the lead shielding box is the random coincidence events  
between \Mo~\twonubb~decay and \g s from \Pb[210], which are enclosed in the 16 tons of the lead shield box.
For the lead shielding box,
the JR Goslar lead bricks are purchased with a \Pb[210]~certification 
and measured activities of \U~and \Th~by the ICP-MS are listed in Table \ref{bkglead}.
As expected, no events are found in the ROI for 2845 and 352 years of exposure for \U~and\Th, respectively.
For a \Pb[210] source, 3.3 \tmEm{7} \g~events are expected in the 35 crystals.  
Considering the \Pb[210]~ activity of \sm59 Bq/kg in the outer lead shield,  2.6 \tmE{4}~\g~events were expected a day and 
the estimated random coincidence rate is 3.4 \tmEm{6} \UNIT.
The total random coincidence rate of \Mo~\twonubb~decay for \Pb[210], \U, and \Th~from the outer lead shield is 3.6 \tmEm{6} \UNIT.

\subsection{Backgrounds from rock material surrounding underground laboratory}
Similar to the lead shielding box, the random coincidence effect of \Mo~\twonubb~decay and \g s from the rock 
are investigated with reported concentrations of \U~and \Th~in rocks at the Y2L \cite{Lee:2011}.
The concentration of \K~is 2.44 ppm, calculated with the natural abundance of K in rocks and natural abundance ratio of \K.
As expected, no events are found in the ROI and 
the estimated upper limits (90$\%$ C.L.) are 3.3 \tmEm{7} and 3.7 \tmEm{7} \UNIT.
The total random coincidence rate of \Mo~\twonubb~decay with \g-rays from \U, \Th, and \K~is 9.2 \tmEm{5} \UNIT.

\subsection{Other backgrounds}

\begin{itemize}

   \item The sandwich panel and cement floor are ones of candidates  with relatively high radioactive sources and huge material mass.
   Concentrations in samples of laboratory wall, called sandwich panel, and a cement floor was measured by the ICP-MS and listed in Table\ref{bklabenv}.
   Similar to the effect of rock and the lead shielding box, only \g s make hits on crystals.
   For instance, for the \Th~sources from sandwich panel and cement floor, 
   only 7.4 \tmEm{7}\% events and 9.6 \tmEm{7}\% events appeared, respectively, and no events are in the ROI.
   The estimated random coincidence rate of \Mo~\twonubb~decays is in the order of \Em{6}~\UNIT.
 \begin{table}[!htb]
\begin{center}
\caption{Concentrations in materials from laboratory environment}
\label{bklabenv}
\begin{tabular}{c|c|c|c}\hline
& \U & \Th & \K \\ \hline \hline
sandwich panel & 1.51 ppm & 1.25 ppm & 202 ppt   \\ \hline
cement floor	& 2.14 ppm & 8.57 ppm & 774 ppt  \\ \hline
\end{tabular}
\end{center}
\end{table}

   \item  Other sources of background (cosmogenic $^{88}$Y, residual $^{48}$Ca in the CMO crystals, 
   and \Bi[214]~ in the copper) are not expected to contribute significantly to the background  near \Mo~\znbb~decay signal region. 
   Nevertheless, they will be considered in the future.
   
\end{itemize} 

\section{Results}
The backgrounds estimates of the AMoRE-I for possible radioactive sources are summarized in Table \ref{bkgAMoREI}. 
The most dominant background source is internal backgrounds in CMO crystals.
After applying rejection cuts such as \a-tagging method and PSD, \b-like event rate from CMO internal background is 
reduced to \lt5.4 \tmEm{5} \UNIT~and \lt8.3 \tmEm{4} \UNIT~for \U~and \Th~decay chains, respectively.
Vikuiti reflectors and CMO supporting copper frames are the next dominant sources to crystals.
After \b-\a~event rejection and \Tl~\b~rejection methods are applied, the event rate 
is reduced to \lt1.1 \tmEm{4} \UNIT~and \lt1.9 \tmEm{4} \UNIT~for \U~and \Th, respectively, in the Vikuiti reflector. 
The event rate with the rejection method in the CMO supporting copper frame is \lt2.3 \tmEm{6} \UNIT~and \lt 2.6 \tmEm{4} \UNIT,
for  \U~and \Th, respectively. 
In the end, the total expected rate in the ROI due to direct background sources is \lt1.5 \tmEm{3} \UNIT. 

\begin{table}[!htb]
  \caption{Summary of \a~and \b-decay-induced (\b-like) backgrounds in major components estimated with measurements and simulation.
99.9999\% alpha rejection and 90\% \b-\a~rejection are assumed. }
\label{bkgAMoREI}
    \begin{threeparttable}
\begin{tabular}{c|c|c|c|c|c|c} \hline
\multirow{5}{*}{ } & \multirow{4}{*}{Isotopes}  & \multirow{2}{*}{Simulated} &\multicolumn{4}{ c }{Backgrounds in ROI} \\ 
Background & &  \multirow{2}{*}{Time} &\multicolumn{4}{ c  }{[\tmEm{3} cnt/keV/kg/yr]} \\ \cline{4-7}
sources & & \multirow{2}{*}{[Years]} & \a  & \b/\g & \b-\a & events \\
& &   & events & events & pileup & with cuts \\ \hline \hline
  \multirow{4}{*}{Internal} & \Pb[210] & 44 & 17 & - & - & - \\
  \multirow{4}{*}{CMO}& \U & 34 & 4.5 & - &  - & - \\ 
 		& \Ra[226] 	& 520 & 0.81 & 0.036 & 0.19 & 0.054 \\ 
		& \Th 		& 700 & \lt 0.55 & \lt 31 & \lt 0.15  & \lt 0.83  \\
  		& \U[235] 	& 68 & 5.5 & - & -\\ \hline
Vikuiti 	& \U 		& 2.3\tmE{4} & 11 & \lt0.041 &  \lt0.66 & \lt 0.11 \\
reflector 	& \Th 	& 4.7\tmE{4} & 4.3 & \lt0.17 & \lt0.18 & \lt0.19 \\ \hline
CMO supporting & \U & 9.9\tmE{3} & 0.013 & \lt0.0022 & \lt0.0013\tnote{*} & \lt0.0023  \\
copper frame & \Th & 5.7\tmE{3} & 0.020 & \lt0.26 & \lt0.0008 & \lt0.26 \\ \hline
SC& \U & 2.6\tmE{4} & - & \lt0.013 & & \lt0.013 \\ 
lead shield & \Th & 8.4\tmE{4} & - & \lt0.0022 & & \lt0.0022 \\ \hline 
 Inner  & \U & 1.5\tmE{3} & - & \lt0.0074\tnote{*} & & \lt0.0074 \\ 
 lead shield& \Th & 1.2\tmE{3} & - & \lt0.0092\tnote{*} & & \lt0.0092  \\ \hline 
Cu plate & \U & 8.7\tmE{3} & - & \lt0.0013\tnote{*} & & \lt0.0013 \\ 
under lead & \Th &  2.5\tmE{4} & - & \lt0.0023 & & \lt0.0023 \\ \hline
G10  	& \U & 2.4\tmE{4} & - & 0.0026 & & 0.0026 \\ 
fiberglass tubes& \Th &  2.5\tmE{4} & - & 0.0064 & & 0.0064 \\ \hline
Stainless  & \U & 5.3\tmE{6} & - & \lt 0.0049 & & \lt 0.0049 \\ 
steel 	& \Th &  2.6\tmE{6} & - & \lt 0.0057 & & \lt 0.0057 \\ \hline \hline
Total 	& & & \lt44 & \lt32 & \lt 1.2& \lt 1.5 \\ \hline
      \end{tabular}
      \begin{tablenotes}
        \footnotesize
        \item[*] 90\% C.L. with zero-entry
      \end{tablenotes}
    \end{threeparttable}
\end{table}

Estimated random coincidence rates listed in Table \ref{RCbkgAMoREI}
are random coincidence rate of two \Mo~\twonubb~decays, 
as well as that of \Mo~\twonubb~decays with other radioactive background sources from possible background sources,
including CMO crystals, Vikuiti reflector, CMO supporting copper frames, G10 fiberglass tubes, 
outer lead shield, rock shell, etc.
The most dominant source for the random coincidence rate is 
two \Mo~\twonubb~decays inside the CMO crystals, 1.2 \tmEm{4} \UNIT.
The next dominant background source for the random coincidence rate with \Mo~\twonubb~decay is \g~from rock shell,
due to its huge mass.
The total estimated random coincidence rate of \Mo~\twonubb~decay with backgrounds is \lt 2.3 \tmEm{4} \UNIT.

\begin{table}[!htb]
\begin{center}
\caption{Backgrounds due to random coincidence with \Mo~\twonubb~decay}
\label{RCbkgAMoREI}
\begin{tabular}{c|c|c} \hline
 \multirow{2}{*}{Material} & \multirow{2}{*}{Sources}  & Random coincidence rate\\ 
 & & [\tmEm{3} cnt/keV/kg/yr] \\
\hline \hline
\multirow{3}{*}{Internal CMO} & two \Mo~\twonubb~decays & 0.12 \\
                              & \Pb[210],\Ra,\Th,\K,\U[235]  & \lt 0.0078  \\
  & two radioactive sources   & \lt 2.2 \tmEm{5} \\ \hline
Vikuiti reflector &  \U, \Th & \lt 1.1\tmEm{5} \\ \hline 
CMO supporting & \multirow{2}{*}{\U, \Th}  &  \multirow{2}{*}{\lt 3.1 \tmEm{5}} \\
copper frame      &  &   \\ \hline
SC lead shield& \Pb[210], \U, \Th & \lt5.8\tmEm{4}  \\ \hline
 Cu Plate &  \U, \Th & \lt 5.8\tmEm{6} \\ \hline 
Inner lead shield & \Pb[210], \U, \Th & \lt 5.0\tmEm{6} \\  \hline 
G10 fiberglass tubes &  \U, \Th & 4.1\tmEm{4}  \\ \hline 
Outer lead shield & \Pb[210], \U, \Th & \lt 0.0036 \\ \hline
Cement floor  & 2.61 MeV \g~(\Th) & 0.0017 \\ \hline
Sandwich panel  & 2.61 MeV \g~(\Th) & 0.0016 \\ \hline
\multirow{2}{*}{\g~from Rock}& 1.76(\U), 2.61 (\Th) &\multirow{2}{*}{0.092} \\ 
& 1.46 MeV (\K) &  \\  \hline  \hline
Total & & \lt 0.23  \\ \hline
\end{tabular}
\end{center}
\end{table}

\section{Discussion}

The most dominant background source in the ROI is \b s from the \Tl~decay in \Th~chain inside CMO crystals.
In this estimation, the \Th~activity inside crystals, 50 $\mu$Bq/kg,
is originated from the upper limit for crystal growing.
In reality, activities of  \U~and \Th~inside each crystal are different, 
so measured activities of each crystal will be used for future work.

For a future experiment after the AMoRE-I, reducing the effect of \Tl~decay inside crystal is the most important. 
Therefore, purification of CMO power and its growing process studies are ongoing. 
In addition,  methods for improving \a-tagging efficiency for rejecting \Tl~have been studied using a simulation.
When the internal background level is reduced to the order of \Em{4}~\UNIT, 
reducing background effect from radioactive sources from Vikuiti reflector and CMO supporting copper frames should be  considered as well.

\section{Conclusion} 
We simulated possible internal and external background sources in the AMoRE-I experiment configuration
and the estimated total background rate in ROI is \lt 1.5 \tmEm{3} \UNIT.
The estimated background level shows that the AMoRE-I experiment will achieve the aimed level of 2 \tmEm{3} \UNIT.
For the AMoRE-I, the main background source is \b s from \Tl[208]~events inside the crystals and materials nearby crystals.
In order to reduce background rate further, R\&D works for crystal purification and material selection are in progress.

\section*{Acknowledgments}

This research was funded by the Institute for Basic Science (Korea) under project code IBS- R016-D1.


\bibliography{amore10bgsimulation}

\begin{thebibliography}{10}
\expandafter\ifx\csname url\endcsname\relax
  \def\url#1{\texttt{#1}}\fi
\expandafter\ifx\csname urlprefix\endcsname\relax\def\urlprefix{URL }\fi
\expandafter\ifx\csname href\endcsname\relax
  \def\href#1#2{#2} \def\path#1{#1}\fi

\bibitem{Mohapatra:2005wg}
R.~N. Mohapatra, et~al., {Theory of neutrinos: A White paper}, Rept. Prog.
  Phys. 70 (2007) 1757--1867.
\newblock \href {http://arxiv.org/abs/hep-ph/0510213}
  {\path{arXiv:hep-ph/0510213}}, \href
  {http://dx.doi.org/10.1088/0034-4885/70/11/R02}
  {\path{doi:10.1088/0034-4885/70/11/R02}}.

\bibitem{Beringer:1900zz}
J.~Beringer, et~al., {Review of Particle Physics (RPP)}, Phys. Rev. D86 (2012)
  010001.
\newblock \href {http://dx.doi.org/10.1103/PhysRevD.86.010001}
  {\path{doi:10.1103/PhysRevD.86.010001}}.

\bibitem{Giunti:2007ry}
C.~Giunti, C.~W. Kim, {Fundamentals of Neutrino Physics and Astrophysics},
  Oxford, UK: Univ. Pr. (2007) 710 p, 2007.

\bibitem{Bhang:2012gn}
H.~Bhang, et~al., {AMoRE experiment: a search for neutrinoless double beta
  decay of Mo-100 isotope with Ca-40 MoO-100(4) cryogenic scintillation
  detector}, J. Phys. Conf. Ser. 375 (2012) 042023.
\newblock \href {http://dx.doi.org/10.1088/1742-6596/375/1/042023}
  {\path{doi:10.1088/1742-6596/375/1/042023}}.

\bibitem{Alenkov:2015dic}
V.~Alenkov, et~al., {Technical Design Report for the AMoRE $0\nu\beta\beta$
  Decay Search Experiment}\href {http://arxiv.org/abs/1512.05957}
  {\path{arXiv:1512.05957}}.

\bibitem{Artusa:2014wnl}
D.~R. Artusa, et~al., {Exploring the Neutrinoless Double Beta Decay in the
  Inverted Neutrino Hierarchy with Bolometric Detectors}, Eur. Phys. J. C74
  (2014) 3096.
\newblock \href {http://arxiv.org/abs/1404.4469} {\path{arXiv:1404.4469}},
  \href {http://dx.doi.org/10.1140/epjc/s10052-014-3096-8}
  {\path{doi:10.1140/epjc/s10052-014-3096-8}}.

\bibitem{Beeman:2011bg}
J.~W. Beeman, et~al., {A next-generation neutrinoless double beta decay
  experiment based on $ZnMoO_4$ scintillating bolometers}, Phys. Lett. B710
  (2012) 318--323.
\newblock \href {http://arxiv.org/abs/1112.3672} {\path{arXiv:1112.3672}},
  \href {http://dx.doi.org/10.1016/j.physletb.2012.03.009}
  {\path{doi:10.1016/j.physletb.2012.03.009}}.

\bibitem{Chernyak:2014ska}
D.~M. Chernyak, F.~A. Danevich, A.~Giuliani, M.~Mancuso, C.~Nones, E.~Olivieri,
  M.~Tenconi, V.~I. Tretyak, {Rejection of randomly coinciding events in
  ZnMoO$_4$ scintillating bolometers}, Eur. Phys. J. C74 (2014) 2913.
\newblock \href {http://arxiv.org/abs/1404.1231} {\path{arXiv:1404.1231}},
  \href {http://dx.doi.org/10.1140/epjc/s10052-014-2913-4}
  {\path{doi:10.1140/epjc/s10052-014-2913-4}}.

\bibitem{esr}
\href{http://www.3m.com/3M/en_US/company-us/all-3m-products/~/3M-Enhanced-Specular-Reflector-3M-ESR-?N=5002385+8709318+8709341+8710654+8710739+8711017+8712789+8718458+3293061534&rt=rud}{The
  3m enhanced specular reflector film}.
\newline\urlprefix\url{http://www.3m.com/3M/en_US/company-us/all-3m-products/~/3M-Enhanced-Specular-Reflector-3M-ESR-?N=5002385+8709318+8709341+8710654+8710739+8711017+8712789+8718458+3293061534&rt=rud}

\bibitem{Lee:2015tsa}
H.~J. Lee, J.~H. So, C.~S. Kang, G.~B. Kim, S.~R. Kim, J.~H. Lee, M.~K. Lee,
  W.~S. Yoon, Y.~H. Kim, {Development of a scintillation light detector for a
  cryogenic rare-event-search experiment}, Nucl. Instrum. Meth. A784 (2015)
  508--512.
\newblock \href {http://dx.doi.org/10.1016/j.nima.2014.11.050}
  {\path{doi:10.1016/j.nima.2014.11.050}}.

\bibitem{Agostinelli:2002hh}
S.~Agostinelli, et~al., {GEANT4: A Simulation toolkit}, Nucl. Instrum. Meth.
  A506 (2003) 250--303.
\newblock \href {http://dx.doi.org/10.1016/S0168-9002(03)01368-8}
  {\path{doi:10.1016/S0168-9002(03)01368-8}}.

\bibitem{Ponkratenko:2000um}
O.~A. Ponkratenko, V.~I. Tretyak, {\relax Yu}.~G. Zdesenko, {The Event
  generator DECAY4 for simulation of double beta processes and decay of
  radioactive nuclei}, Phys. Atom. Nucl. 63 (2000) 1282--1287, [Yad.
  Fiz.63,1355(2000)].
\newblock \href {http://arxiv.org/abs/nucl-ex/0104018}
  {\path{arXiv:nucl-ex/0104018}}, \href {http://dx.doi.org/10.1134/1.855784}
  {\path{doi:10.1134/1.855784}}.

\bibitem{Alessandria:2012zp}
F.~Alessandria, et~al., {Validation of techniques to mitigate copper surface
  contamination in CUORE}, Astropart. Phys. 45 (2013) 13--22.
\newblock \href {http://arxiv.org/abs/1210.1107} {\path{arXiv:1210.1107}},
  \href {http://dx.doi.org/10.1016/j.astropartphys.2013.02.005}
  {\path{doi:10.1016/j.astropartphys.2013.02.005}}.

\bibitem{Kim:2015pua}
G.~B. Kim, et~al., {A CaMoO4 Crystal Low Temperature Detector for the AMoRE
  Neutrinoless Double Beta Decay Search}, Adv. High Energy Phys. 2015 (2015)
  817530.
\newblock \href {http://dx.doi.org/10.1155/2015/817530}
  {\path{doi:10.1155/2015/817530}}.

\bibitem{gbkimthesis}
G.-B. Kim, A 0$\nu\beta\beta$ search using large scintillating crystal with
  metallic magnetic calorimeter, Ph.D. thesis, Seoul National University
  (2016).

\bibitem{2016JPhCS.718f2050S}
E.~{Sala}, I.~S. {Hahn}, W.~G. {Kang}, G.~W. {Kim}, Y.~D. {Kim}, M.~H. {Lee},
  D.~S. {Leonard}, S.~Y. {Park}, {Development of an underground low background
  instrument for high sensitivity measurements}, Journal of Physics Conference
  Series 718~(6) (2016) 062050.
\newblock \href {http://dx.doi.org/10.1088/1742-6596/718/6/062050}
  {\path{doi:10.1088/1742-6596/718/6/062050}}.

\bibitem{Alessandria:2011vj}
F.~Alessandria, et~al., {CUORE crystal validation runs: results on radioactive
  contamination and extrapolation to CUORE background}, Astropart. Phys. 35
  (2012) 839--849.
\newblock \href {http://arxiv.org/abs/1108.4757} {\path{arXiv:1108.4757}},
  \href {http://dx.doi.org/10.1016/j.astropartphys.2012.02.008}
  {\path{doi:10.1016/j.astropartphys.2012.02.008}}.

\bibitem{Kim:2015prep}
{in preparation}.

\bibitem{Chernyak:2012zz}
D.~M. Chernyak, F.~A. Danevich, A.~Giuliani, E.~Olivieri, M.~Tenconi, V.~I.
  Tretyak, {Random coincidence of 2nu 2beta decay events as a background source
  in bolometric 0nu 2beta decay experiments}, Eur. Phys. J. C72 (2012) 1989.
\newblock \href {http://arxiv.org/abs/1301.4248} {\path{arXiv:1301.4248}},
  \href {http://dx.doi.org/10.1140/epjc/s10052-012-1989-y}
  {\path{doi:10.1140/epjc/s10052-012-1989-y}}.

\bibitem{Belli:2010zzc}
P.~Belli, et~al., {New observation of 2beta 2nu decay of Mo-100 to the 0+(1)
  level of Ru-100 in the ARMONIA experiment}, Nucl. Phys. A846 (2010) 143--156.
\newblock \href {http://dx.doi.org/10.1016/j.nuclphysa.2010.06.010}
  {\path{doi:10.1016/j.nuclphysa.2010.06.010}}.

\bibitem{Blum:1992vr}
D.~Blum, et~al., {Search for gamma-rays following beta beta decay of Mo-100 to
  excited states of Ru-100.}, Phys. Lett. B275 (1992) 506--511.
\newblock \href {http://dx.doi.org/10.1016/0370-2693(92)91624-I}
  {\path{doi:10.1016/0370-2693(92)91624-I}}.

\bibitem{Leonard:2016prep}
in preparation.

\bibitem{Laubenstein:2009}
M.~Laubenstein, G.~Geusser, {Cosmogenic radionuclides in metals as indicator
  for sea level exposure history}, Appl. Rad. Isot. 67 (2009) 750.
\newblock \href {http://dx.doi.org/10.1016/j.apradiso.2009.01.029}
  {\path{doi:10.1016/j.apradiso.2009.01.029}}.

\bibitem{SNO91_15}
A.~R. Smith, D.~L. Hurley,
  \href{http://www.sno.phy.queensu.ca/sno/str/index.html}{Gamma spectrometric
  analyses of materials}, SNO Technical Reports Rad SNO-STR-91-015 (1991).
\newline\urlprefix\url{http://www.sno.phy.queensu.ca/sno/str/index.html}

\bibitem{Lee:2011}
M.~Lee, et~al., {Radon Environment in the Korea Invisible Mass Search
  Experiment and Its Measurement}, J. Kor. Phys. Soc. 58 (2011) 713.
\newblock \href {http://dx.doi.org/10.3938/jkps.58.713}
  {\path{doi:10.3938/jkps.58.713}}.

\end{thebibliography}

\end{document}